\documentclass[conference]{IEEEtran}
\IEEEoverridecommandlockouts
\usepackage{cite}
\usepackage{amsmath,amssymb,amsfonts}
\usepackage{algorithmic}
\usepackage{graphicx}
\usepackage{textcomp}
\usepackage{xcolor}
\usepackage{multirow}
\usepackage[super]{nth}
\def\BibTeX{{\rm B\kern-.05em{\sc i\kern-.025em b}\kern-.08em
    T\kern-.1667em\lower.7ex\hbox{E}\kern-.125emX}}

\makeatletter
\newcommand{\linebreakand}{%
  \end{@IEEEauthorhalign}
  \hfill\mbox{}\par
  \mbox{}\hfill\begin{@IEEEauthorhalign}
}
\makeatother

\newcommand{\blk}{\color{black}}

\begin{document}


\title{Towards Neural Decoding of Imagined Speech based on Spoken Speech
\footnote{{\thanks{This work was partly supported by Institute for Information \& Communications Technology Planning \& Evaluation (IITP) grant funded by the Korea government (MSIT) (No.2021-0-02068, Artificial Intelligence Innovation Hub; No. 2017-0-00451, Development of BCI based Brain and Cognitive Computing Technology for Recognizing User’s Intentions using Deep Learning).}
}}
}




\author{\IEEEauthorblockN{Seo-Hyun Lee}
\IEEEauthorblockA{\textit{Dept. Brain and Cognitive Engineering} \\
\textit{Korea University} \\
Seoul, Republic of Korea \\
seohyunlee@korea.ac.kr} \\

\and

\IEEEauthorblockN{Young-Eun Lee}
\IEEEauthorblockA{\textit{Dept. Brain and Cognitive Engineering} \\
\textit{Korea University} \\
Seoul, Republic of Korea \\
ye\_lee@korea.ac.kr} \\

\linebreakand 

\IEEEauthorblockN{Soowon Kim}
\IEEEauthorblockA{\textit{Dept. Artificial Intelligence} \\
\textit{Korea University} \\
Seoul, Republic of Korea \\
soowon\_kim@korea.ac.kr} \\

\and 

\IEEEauthorblockN{Byung-Kwan Ko}
\IEEEauthorblockA{\textit{Dept. Artificial Intelligence} \\
\textit{Korea University} \\
Seoul, Republic of Korea \\
leaderbk525@korea.ac.kr} \\

\and

\IEEEauthorblockN{Seong-Whan Lee}
\IEEEauthorblockA{\textit{Dept. Artificial Intelligence} \\
\textit{Korea University} \\
Seoul, Republic of Korea \\
sw.lee@korea.ac.kr}
}

\maketitle



\begin{abstract}
\blk Decoding imagined speech from human brain signals is a challenging and important issue that may enable human communication via brain signals. While imagined speech can be the paradigm for silent communication via brain signals, it is always hard to collect enough stable data to train the decoding model. Meanwhile, spoken speech data is relatively easy and to obtain, implying the significance of utilizing spoken speech brain signals to decode imagined speech. In this paper, we performed a preliminary analysis to find out whether if it would be possible to utilize spoken speech electroencephalography data to decode imagined speech, by simply applying the pre-trained model trained with spoken speech brain signals to decode imagined speech. While the classification performance of imagined speech data solely used to train and validation was 30.5 ± 4.9 \%, the transferred performance of spoken speech based classifier to imagined speech data displayed average accuracy of 26.8 ± 2.0 \% which did not have statistically significant difference compared to the imagined speech based classifier (p = 0.0983, chi-square = 4.64). For more comprehensive analysis, we compared the result with the visual imagery dataset, which would naturally be less related to spoken speech compared to the imagined speech. As a result, visual imagery have shown solely trained performance of 31.8 ± 4.1 \% and transferred performance of 26.3 ± 2.4 \% which had shown statistically significant difference between each other (p = 0.022, chi-square = 7.64). Our results imply the potential of applying spoken speech to decode imagined speech, as well as their underlying common features.
\end{abstract}

\begin{small}
\end{small}

\begin{figure*}[t]
\centering
    \includegraphics[width=\textwidth]{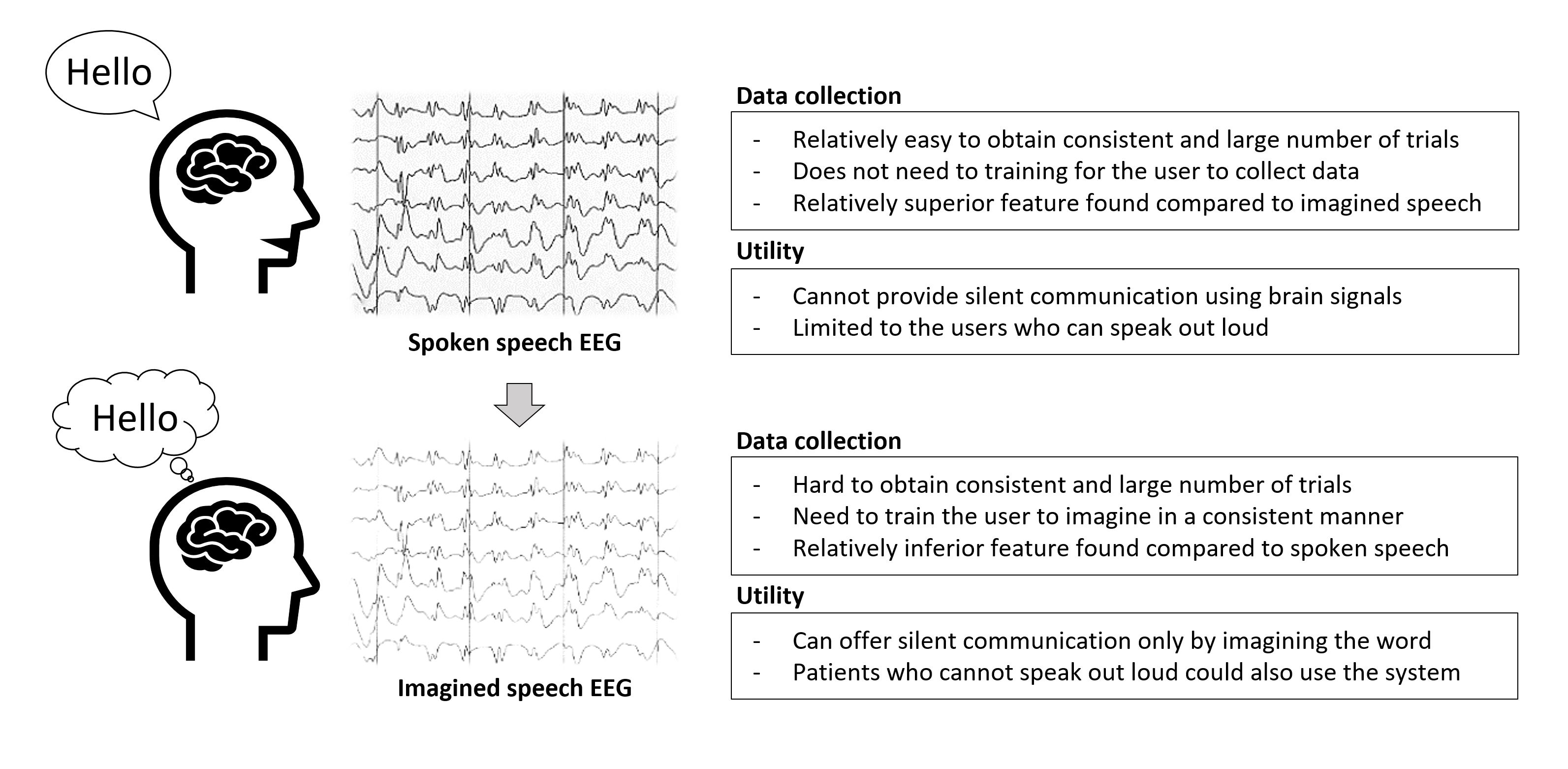}
    \caption{Comparison of spoken speech and imagined speech based BCI communication system.}
    \label{figure1}
\end{figure*}

\section{INTRODUCTION}
Brain-computer interface (BCI) is a technology of converting user's intention to an external output or action via decoding brain signals. It accompanies the imagery of the user, and the process of decoding user's intention from the brain signals. There are specific patterns of brain signals that the BCI system aims to decode, which consists of external stimulus or user's spontaneous imagery including the user's intention \cite{lee2019towards}. Exogenous BCI paradigms, such as event-related potential or steady-state evoked potential, have been actively invested, since it have shown effectiveness in conveying user's intention in a relatively high speed and accuracy \cite{won2017motion, lee2019connectivity, lee2018high}. However, current research stream on BCI is highly focusing on the endogenous paradigms, such as motor imagery \cite{jeong2020decoding, suk2014predicting, jeong2020brain}, imagined speech\cite{lee2020neural}, or visual imagery\cite{sousa2017pure, lee2020spatio}, since they do not require external stimuli, therefore, may be a more convenient way to convey user's intention directly \cite{lee2022toward}.

However, endogenous BCI paradigms yet hold limitations of low decoding performance, and inferior degree-of-freedom (DOI) \cite{sousa2017pure}. In addition, it is relatively hard to acquire consistent brain signal data per each class, since it is not a stimulus-driven brain signals \cite{lee2019towards}. Some users are known to be inefficient in utilizing endogenous BCI paradigms, therefore, may need the process of training the user beforehand. While endogenous BCI paradigms are strong in convenience and intuitiveness, limited amount of high-quality data and the lack of strong features have always been a challenging issue to be addressed.

Imagined speech is an emerging endogenous paradigm for intuitive BCI communication, which refers to the internal imagery of speech, without emitting vocal sound nor moving the mouth \cite{lee2020neural}. It is easy to expand the DOI of imagined speech since there are various words or sentences to be decoded, therefore, may be a strong BCI paradigm that can convey unconstrained intention of the user. However, it is relatively hard to collect imagined speech data compared to the exogenous paradigms or other paradigms, since it is hard to collect consistent imagery data to train the model from the user. Also, the main limitation of the endogenous paradigms are that the data collector cannot ensure if the user consistently imagined the exact right thing, or just had thought of something else, since we cannot check the imagined ground-truth by vision nor hearing.

Spoken speech refers to the natural speech that we use in the everyday life \cite{meng2022evidence}. It is known that imagined speech brain signals resemble the features of spoken speech brain signals in some portion, therefore, holds potential to utilize spoken speech data to improve or enhance the imagined speech decoding performance\cite{timothee2022imagined, miguel2021real}. Unlike imagined speech, spoken speech data is relatively easy to be acquired, and is able to check whether the user performed the speech correctly, therefore, may be better to robustly train the decoding model. Although spoken speech holds strength in terms of data collection, decoding imagined speech is still the most crucial point in the field of BCI, since the first aim for BCI systems is to help patients who cannot move or talk \cite{zhang2017hybrid}.

In this paper, we explored the possibility of utilizing the spoken speech brain signal data to decode imagined speech electroencephalography (EEG). This is a preliminary study of simply applying the spoken speech-based trained model to the imagined speech EEG data, to find out the potential of transferring the robust model trained with spoken speech to imagined speech data which has relatively weak features. We first tested the imagined speech EEG with the spoken speech based trained model. Also we compared the imagined speech result in the same method with the visual imagery dataset to find out the difference between the two paradigms.

\section{Materials and methods}

\subsection{Overall framework}
As shown in the Figure~\ref{figure1}, spoken speech data is relatively easy to obtain in large number of trials and with consistency, as it is a natural speech that most people perform in everyday life. Additionally, training users is not necessary, and distinct brain features have been identified in comparison to imagined speech. However, it cannot coincide with the imagined speech, since the usage of spoken speech is only limited to the people who can speak out loud, and therefore, cannot establish silent communication system that operates only with brain signals. Therefore, our preliminary task was to utilize the strength of both spoken speech and imagined speech, to further transfer the spoken speech based pre-trained model to the imagined speech EEG data.


\begin{table}[!htbp]
    \centering
    \caption{Decoding performance of imagined speech using imagined speech and spoken speech based pre-trained model}
\begin{tabular}{|c|c|c|c|}
\hline
                   & \textbf{\begin{tabular}[c]{@{}c@{}}Imagined speech\\ 10-fold \\ cross validation\end{tabular}} & \textbf{\begin{tabular}[c]{@{}c@{}}Spoken speech\\ full trials\end{tabular}} & \textbf{\begin{tabular}[c]{@{}c@{}}Spoken speech\\ few trials\end{tabular}} \\ \hline
\textbf{Subject 1} & 30.0                                                                                           & 23.4                                                                         & 25.0                                                                        \\ \hline
\textbf{Subject 2} & 35.6                                                                                           & 26.6                                                                         & 25.2                                                                        \\ \hline
\textbf{Subject 3} & 38.1                                                                                           & 28.6                                                                         & 29.1                                                                        \\ \hline
\textbf{Subject 4} & 28.7                                                                                           & 29.5                                                                         & 23.2                                                                        \\ \hline
\textbf{Subject 5} & 23.7                                                                                           & 26.4                                                                         & 31.4                                                                        \\ \hline
\textbf{Subject 6} & 27.3                                                                                           & 25.9                                                                         & 34.3                                                                        \\ \hline
\textbf{Subject 7} & 30.4                                                                                           & 26.8                                                                         & 28.0                                                                        \\ \hline
\textbf{AVG.}      & 30.5                                                                                           & 26.8                                                                         & 28.0                                                                        \\ \hline
\textbf{STD.}      & 4.9                                                                                            & 2.0                                                                          & 3.9                                                                         \\ \hline
\end{tabular}
\end{table}

\begin{table}[!htbp]
    \centering
    \caption{Decoding performance of visual imagery using visual imagery and spoken speech based pre-trained model}
\begin{tabular}{|c|c|c|c|}
\hline
                   & \textbf{\begin{tabular}[c]{@{}c@{}} Visual imagery\\ 10-fold \\ cross validation\end{tabular}} & \textbf{\begin{tabular}[c]{@{}c@{}}Spoken speech\\ full trials\end{tabular}} & \textbf{\begin{tabular}[c]{@{}c@{}}Spoken speech\\ few trials\end{tabular}} \\ \hline
\textbf{Subject 1} & 33.9                                                                                           & 25.0                                                                         & 24.5                                                                        \\ \hline
\textbf{Subject 2} & 25.3                                                                                           & 24.1                                                                         & 25.9                                                                        \\ \hline
\textbf{Subject 3} & 33.3                                                                                           & 27.5                                                                         & 31.8                                                                        \\ \hline
\textbf{Subject 4} & 33.4                                                                                           & 29.1                                                                         & 26.6                                                                        \\ \hline
\textbf{Subject 5} & 35.2                                                                                           & 22.7                                                                         & 23.2                                                                        \\ \hline
\textbf{Subject 6} & 26.4                                                                                           & 28.0                                                                         & 24.3                                                                        \\ \hline
\textbf{Subject 7} & 35.1                                                                                           & 27.7                                                                         & 22.7                                                                        \\ \hline
\textbf{AVG.}      & 31.8                                                                                           & 26.3                                                                         & 25.6                                                                        \\ \hline
\textbf{STD.}      & 4.1                                                                                            & 2.4                                                                          & 3.1                                                                         \\ \hline
\end{tabular}
\end{table}

\begin{figure}[t]
\centering
    \includegraphics[width=\linewidth]{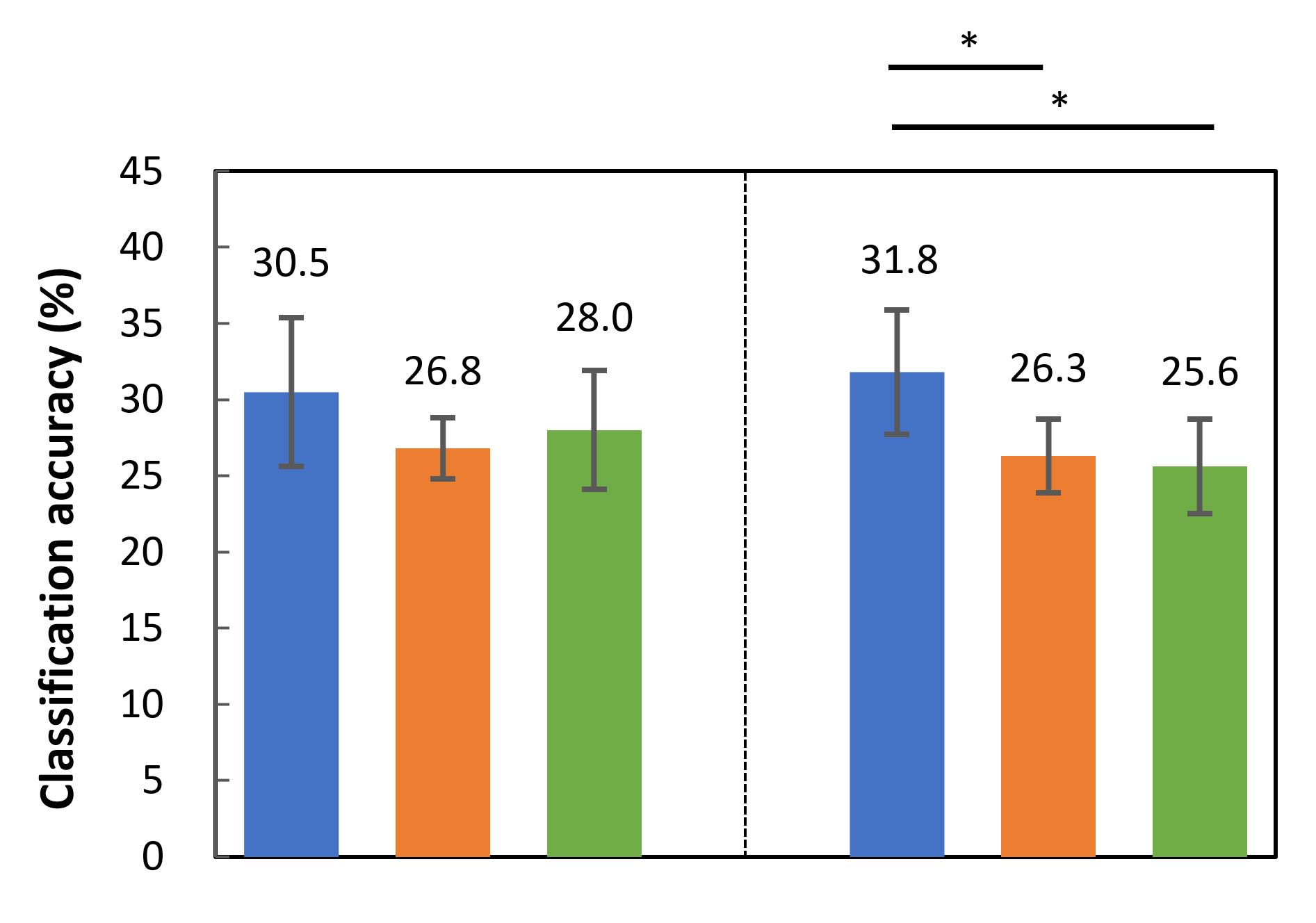}
    \caption{Comparison of the classification performance of imagined speech(left) and visual imagery(right).The blue bars represent the mean performance of 10-fold cross validation results. The orange bars represent the mean performance using full spoken speech trials based pre-trained model and the green bars represents the mean performance of few spoken trials applied model.}
    \label{figure2_2}
\end{figure}


\subsection{Data Acquisition}
\subsubsection{Participants}
Spoken speech, imagined speech, and visual imagery EEG dataset of 7 subjects were used in this study. The dataset was acquired from the previous studies \cite{lee2019towards, lee2019eeg, jeong20222020, lee2020neural}. The study was carried out in accordance with the Declaration of Helsinki. The experimental protocols were reviewed and approved by the Institutional Review Board at Korea University [KUIRB-2019-0143-01] and all subjects signed informed consent.

\subsubsection{Experimental Setup}
64-channel EEG cap with active electrodes placement following the international 10-10 system were used for the recording. Reference and ground electrodes were set to FCz and FPz channels, respectively. EEG signals were recorded via Brain Vision/Recorder (BrainProduct GmbH, Germany) and operated by MatLab 2018a software.

\subsubsection{Experimental Paradigm}
Experimental paradigms are explained in detail in the previous studies \cite{lee2019towards, lee2019eeg, lee2020neural}. The dataset of spoken speech, imagined speech, and visual imagery consists of the same words/phrases from the same participants. In this paper, 5-class words were selected from the dataset to test the transfer scenario as a preliminary study.

\subsection{EEG Data Classification}
\subsubsection{Imagined speech decoding}
10-fold cross validation was performed using 90 \% of randomly selected imagined speech data as a training set and the remaining 10 \% as a test set. This was set as the baseline performance to compare with the performance of spoken speech based transferred classifier. Support vector machine (SVM) classifier was trained with common spatial pattern (CSP) feature in all three modes of classification, including the following subsections.

\subsubsection{Imagined speech decoding with spoken speech based pre-trained model}
The model trained with spoken speech dataset was transferred to the imagined speech data. Weights for the CSP filters were first trained with spoken speech EEG and applied to the imagined speech data. We tested in two different sets of spoken speech dataset, as one was a model trained with full spoken speech trials and another was the model trained with only a small number of spoken speech trials (10 trials per class).

\subsubsection{Visual imagery decoding with spoken speech based pre-trained model}
Comparing the performance with visual imagery dataset was performed to confirm the viability of transferring spoken speech brain signals to imagined speech brain signals. As same as the case of imagined speech, the model trained with spoken speech dataset was transferred to test the visual imagery data. Weights for the CSP filters were first trained with spoken speech EEG and applied to the visual imagery data.

\subsection{Statistical Analysis}
For the statistical analysis, we performed Kruskal–Wallis test to compare the classification accuracy of the baseline performance of non-transferred 10-fold cross validation, spoken speech based transferred result using full trials, and few trials. Non-parametric bootstrap analysis was applied as a post-hoc analysis. Significance level was set to 0.05.


\section{Results and Discussion}
\subsection{Imagined Speech Decoding}
As shown in the Table 1, the averaged classification performance of imagined speech data solely used to train and test was 30.5 ± 4.9 \%, and the transferred performance of spoken speech based classifier to imagined speech data was 26.8 ± 2.0 \%. The spoken speech based transferred result trained with only few spoken speech trials was 28.0 ± 3.9 \%. Based on the statistical analysis, there was no significant difference found between the imagined speech 10-fold cross validation result with the spoken speech based transferred result (p = 0.0983, chi-square = 4.64). The result exhibits comparable performance of the transferred model, which implies the potential of applying spoken speech dataset to decode imagined speech. Since spoken speech data is much simple and easier to acquire, it would be more efficient to train and transfer models using large spoken speech trials, provided that comparable results can be achieved.


\subsection{Visual Imagery Decoding}
For more comprehensive analysis, we compared the result with the visual imagery dataset, which would naturally be less related to the spoken speech compared to the imagined speech \cite{lee2021functional}. As shown in the Table 2, visual imagery have shown solely trained performance of 31.8 ± 4.1 \% and transferred performance of 26.3 ± 2.4 \% which had significant statistical difference between each other (p = 0.022, chi-square = 7.64). Since there was statistically significant difference only for the case of visual imagery (Figure 2), this result confirms that there may be common features between the two speech-related paradigms (imagined speech and spoken speech).

\section{Conclusion}
Our result implies the potential of applying spoken speech to decode imagined speech, as well as their underlying common features. Our experiments revealed that the imagined speech may be more related to the spoken speech than the visual imagery. Conclusively, decoding imagined speech via spoken speech-based models may hold great promise in assisting individuals with physical or speech impairments and represents a valuable area of research as a state-of-the-art communication technology that can produce actual vocalizations through the imagined speech. Since we have just simply transferred the pre-trained model to a different dataset, such method as fine tuning would further improve the result. For the future work, we would apply advanced deep learning model to pre-train the model with spoken speech dataset, and fine tune the model using small amount of imagined speech data for better results \cite{thung2018conversion, kim2019subject, kwon2019subject, lee2022eeg}.


\bibliographystyle{IEEEtran}

\bibliography{REFERENCE}

\end{document}